# EDAP: An Efficient Data-Gathering Protocol for Wireless Sensor Networks

## Arash Nasiri Eghbali, Nastooh Taheri Javan* and Mehdi Dehghan

Ad Hoc Network laboratory
Computer Engineering and Information Technology Department
Amirkabir University of Technology (Tehran Polytechnic), Tehran, Iran
Email: nasiri@aut.ac.ir
Email: nastooh@aut.ac.ir
Email: dehghan@aut.ac.ir
* Corresponding author

**Abstract:** Directed Diffusion (DD), uses mechanisms such as data aggregation and in-network processing to suppress the additional data overhead however there is no guarantee that paths from nearby sources join after a few hops. Also for path construction, each source floods a distinct exploratory data (ED) packet through the network which is not necessary for nearby sources.

In this paper an efficient data-gathering protocol (EDAP) is proposed to address the problems of late-aggregation and distinct ED-flooding through using a virtual sink (VS) node nearby the sources which plays the role of sink node and broadcasts local interest messages and routed gathered data toward destination.

Also in this protocol, multiple paths are constructed between virtual sink and the sink node which leads to load-balancing and increase in the life-time of the network nodes. Our routing protocol can produce more than one disjoint or braided path and spread the gathered data properly between constructed paths according to their length and remaining energy. Simulation results shows that in this way a significant amount of energy can be saved and the network lifetime will be increased considerably (up to 75% better than DD).

**Biographical notes:** Arash Nasiri Eghbali was born in Iran. He obtained his Bachelor's Degree in Electrical Engineering at Tabriz University and the Master's Degree in Computer Networks at Amirkabir University of Technology (Tehran Polytechnic) in 2007. He has been involved in research activities including design and implementation of routing, data-gathering and time synchronization protocols in wireless sensor networks since 2004. His research interests include Network Architecture and Protocols, Distributed Computing, Mobile AD-HOC Networks and Wireless Sensor Networks.

Nastooh Taheri Javan was born in Iran. He obtained his Bachelor's Degree in Computer/Software Engineering at Azad University and the Master's Degree in Computer Architecture Engineering at Amirkabir University of Technology (Tehran Polytechnic) in 2007. He has been involved in research activities including design routing Protocols, Security and fault tolerance in Mobile AD-HOC Networks and data-gathering, Energy Consumption and Routing Protocols in wireless sensor networks since 2004. His research interests include Network Architecture, Network Security, Mobile AD-HOC Networks and Wireless Sensor Networks.

Mehdi Dehghan is an associate professor in the Department of Computer Engineering at the Amirkabir University of Technology in Tehran. He received his B.Sc. in Computer engineering from Iran University of Science and Technology (IUST), Tehran, Iran in 1992, and his M.Sc. and PhD from Amirkabir University of Technology (AUT), Tehran, Iran in 1995, and 2001, respectively. Since 1995, he has been a research scientist at Iran Telecommunication Research Center (ITRC) working in the area of Quality of Service provisioning and Network Management. He joined Computer Engineering Department of Amirkabir University of Technology in 2004. His research interests include High Speed Networks, Network Management, Mobile AD-HOC Networks and Fault Tolerant Computing.

# 1 Introduction

Wireless sensor networks (WSNs) consist of a number of sensor nodes, scattered in the environment to sense special events such as searching a mobile target or measuring the amount of radioactive radiation in a specified area. These nodes are usually identical with a limited amount of energy. Instead of unique addresses, they are identified by the information gathered by them. They are mostly implemented densely and thus a number of nodes are triggered by a single event. In such situations, each node starts to send the gathered information toward the sink which is usually located far from senders.

Directed Diffusion (DD) is a data-centric routing protocol that uses only local interaction between neighbor nodes (Intanagonwiwat et al. 2002). In DD, attribute-value pairs are used for describing information and data. This algorithm in its basic form has two phases. In the first phase, the sink nodes floods a request packet called "interest" which is consist of the desired attribute-value pairs. When this packet reaches a source node that has the requested information (second phase), It floods an "exploratory data" (ED) packet through the network. When this packet reaches the sink node, it will send a "positive reinforcement" packet toward source node. This packet is being forwarded through the path traversed by the ED packet. In this way, a bi-directional path is constructed between two nodes. Afterwards data packets will be sent through reinforced paths. This algorithm is also called two-phase-pull (TPP) algorithm (Heidemann et al. 2003). Also by assuming the connections to be bi-directional (which in most cases is not true), One-Phase-Pull (OPP) algorithm can be used. In this approach, data packets are sent immediately after the interest packet reaches the source node. Therefore in OOP algorithm the cost of ED packet flooding will be omitted.

To provide connection between sinks and sources, this work relies on low-rate flooding of events, enabling local re-routing whenever the nodes in the primary path, have failed due to energy consumption. In sensor networks, where energy efficiency is of paramount importance, such flooding can adversely impact the lifetime of network.

In DD, data aggregation and in-network processing approaches has been introduced to suppress the additional data overhead. In This protocol, the paths from the adjacent sources often join together after a few hops but there are no guarantee for such combination. In cases where the sensed event is spread geographically, the probability of such combination is reduced. Another problem of DD which arises in presence of many source nodes near a single event is the mechanism used in this protocol for path construction between sources and sinks. By receiving an interest, each source floods an exploratory data (ED). When this packet reaches the destination, the path traversed, is reinforced by the sink and used for later data transmissions. The procedure is repeated for each source separately and thus, a significant amount of network energy is consumed.

The EDAP protocol is proposed to address the problems of late-aggregation and separate ED- flooding. In our local clustering protocol, early aggregation can be achieved by using a virtual sink (VS) near the sources which plays the role of sink node and broadcasts local interest messages. Therefore data packets are sent preliminary toward VS node. VS node undertakes the responsibility of sending the data packets toward destination. Hence there is no need for different sources in a single cluster to broadcast separate ED messages. This operation is done only once by the VS node. So the routing protocol overhead is reduced significantly using our clustering method.

Another problem is the mechanism used in DD for routing selection which mostly, leads to the selection of the shortest path between sinks and sources. In this case the nodes in the shortest path will fail after a short period of time due to lack of energy. This problem is intensified when the nearest route to the shortest path, is selected after path refresh period which is quite probable. In this situation, network partitioning will occur along the depleted paths.

To avoid this, in our paper, we propose using multipath routing algorithms to increase load-balancing between network nodes during forwarding data packets, between VS and sink nodes. In this way, we can reduce the rate of interest packet flooding by increasing the lifetime of connection. We consider a proactive approach to construct multiple paths between two nodes and use node-disjoint multi-paths where the alternate paths do not intersect each other. The disjoint property has better performance for our load-balancing technique but reduces the number of paths, created by the algorithm. Also if a single node in a disjoint path fails, other nodes will be left unused until next refreshment period. In sensor networks, designing such algorithms has been proved to be a difficult task, due to their data-centric routing with localized path setup as indicated in Ganesan et al. (2001).

Limited forward improvement (LFI) improvement method is used in this paper to improve the probability of construction of multiple paths. We introduce the premier packet problem where the first flooded packet, dominates most of the nodes between source and destination and prevents our proactive algorithm to construct multiple paths. Using limited forward improvement, each node would selectively send the exploratory data to nodes nearer to destination instead of broadcasting them. In this way, we can achieve a significant reduction in the overhead imposed by Directed Diffusion to broadcast the exploratory data packets all through the network. The number of constructed paths would also grow using this improvement method. This method can be used instead of two phase pull directed diffusion which uses a network wide flooding for each source node and one phase pull directed diffusion where all the connections are assumed to be bi-directional.

Each of these algorithms separately has its own weak points. When Using on demand clustering protocol, all paths from different sources will merge after a few hops and all of them are forwarded by virtual sink node toward the sink, using a single path. As mentioned earlier this may cause the network to be partitioned. Also multi-path routing algorithm constructs disjoint paths between sources and sink nodes and prevents aggregation. The Efficient Data Gathering Protocol (EDAP) uses both algorithms to gain the benefits of each one and bypass their defects. A sample of routing scheme in EDAP protocol is showed in figure 1.

The rest of this paper is presented as following: in section 2, we will introduce on-demand clustering protocol and describe the algorithm in detail. In section 3 we explain the algorithm used for multiple paths construction and also introduce limited forward improvement method. We also present our load-balancing mechanism used after multi-path construction phase in this section. In section 4 we will explain the EDAP algorithm and discuss about the problems that we may face during the combination of the on-demand clustering and multipath routing algorithms. The methodology we used for implementing and testing these protocols and simulation results are available in section 5 and a comparison between the original directed diffusion algorithm and our proposed algorithms will be given in this section. Related works are reviewed in section 6 and finally we will conclude the paper and present future works to improve our routing algorithm in section 7.

**Figure 1** Sample EDAP routing scheme

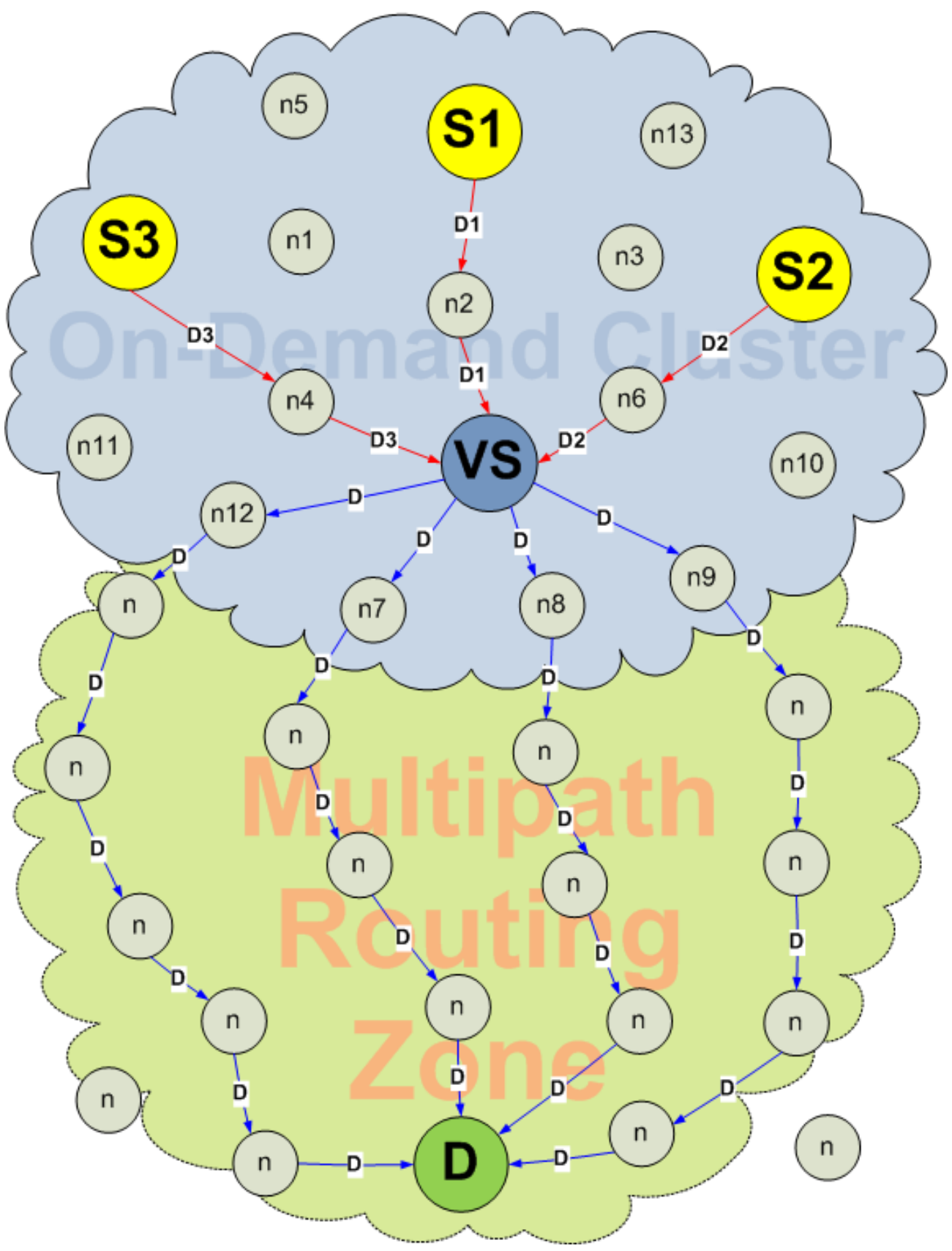

## 2 On Demand Clustering Protocol

On-Demand Clustering Protocol (ODCP) is used for collecting information gathered by nearby sensor nodes (Nasiri Eghbali et al. 2007). Then these data are aggregated and sent toward sink by the cluster head named as virtual sink (VS). In this section the ODCP clustering protocol will be introduced and all the mechanisms used for cluster head selection and routing within a cluster will be described in detail.

This protocol has four phases. First is the virtual sink selection in which a suitable cluster-head is selected among nodes near the sources. In the second phase, selected VS will explore a path to the sink and in the third phase, after a period of time a new virtual sink is selected among neighbors of prior VS. The fourth phase is considered for situations where the sink node is crashed or lacks enough energy to continue forwarding data packets toward the sink.

**Phase I: Virtual Sink Selection** – virtual sink selection is the most important and challenging phase in this protocol. As DD is a localized algorithm and each node has only local information, selecting a single node as virtual sink node between others seems to be a difficult task. The first idea which crosses the mind is to select one of the sources as VS. Problems will occur when sources are not in radio range of each other. Hence in such case, multiple VSs will be selected and we need to choose one of them as VS. In ODPC one of the nodes in original path between source and sink will be selected as VS. Distance of this node from the source (Dsrc) can be selected, considering density of nodes and spatial properties of sensed event. As we usually prefer to increase the life-time of source nodes, this selection seems to be a better one. In our simulations, Dsrc assumed to be at most 2 hops. When a node is selected as cluster-head, it broadcasts an interest message to all nodes in the cluster. For this purpose, limited flooding method is used. The interest packets are tagged by VS. This tag indicates the time-to-live (TTL) of forwarded message and each packet will be ignored after traversing such amount of nodes. The TTL value is decremented in each hop until it becomes zero. Therefore the overhead of such flooding is directly related to the size of cluster which is very small in comparison with network size.

In this step, the number of selected VS nodes is the same as number of sources. This selection is shown in the figure 2.c. Firstly VS1, VS2 and VS3 are selected as virtual sink; so one of these VS nodes should be selected as the final VS. A simple approach for such selection is tagging the VS interest messages by VS selection time. Cluster nodes (including VS nodes) can select VS with the least time-stamp. So the nearest VS to the sink is selected. Overhead of such flooding increases linearly by the number of sources which is not acceptable especially in high density networks.

In figure 2.c the VS selection mechanism has been shown. Here the VS1 node is the VS with the lowest time-stamp. So the packets from VS2 and VS3 are not flooded in cluster. The paths used by the original DD are depicted in figure 2.a and the paths used by ODCP are shown in figure 2.b. In ODCP, we used two-phase-pull algorithm to route packets within the cluster and sources send local exploratory data to the VS and it reinforces them.

Using this method, the nodes having a distance less than Dsrc from the selected VS node, will form a cluster.

**Phase II: Route Discovery** – After VS selection phase, we should construct a path from each source toward VS and a path from VS to the sink. Two approaches are used for path construction within the cluster. First is using One-Phase-Pull (OPP) algorithm within the cluster. In this case sources will send data through the path traversed by local interest messages. As we did not assume the links to be bi-directional, the realistic approach is using Two-Phase-Pull (TPP) algorithm in which sources will broadcast a local ED packet in the cluster and VS node reinforces the constructed path after receiving an ED packet from each source.

Then in order to find a path toward the sink, TPP algorithm is used and VS broadcasts an ED packet. When this packet reaches sink, this node will reinforce the path traversed by the ED packet and a path will be constructed from VS toward the sink.

**Figure 2** The On-Demand Clustering Protocol

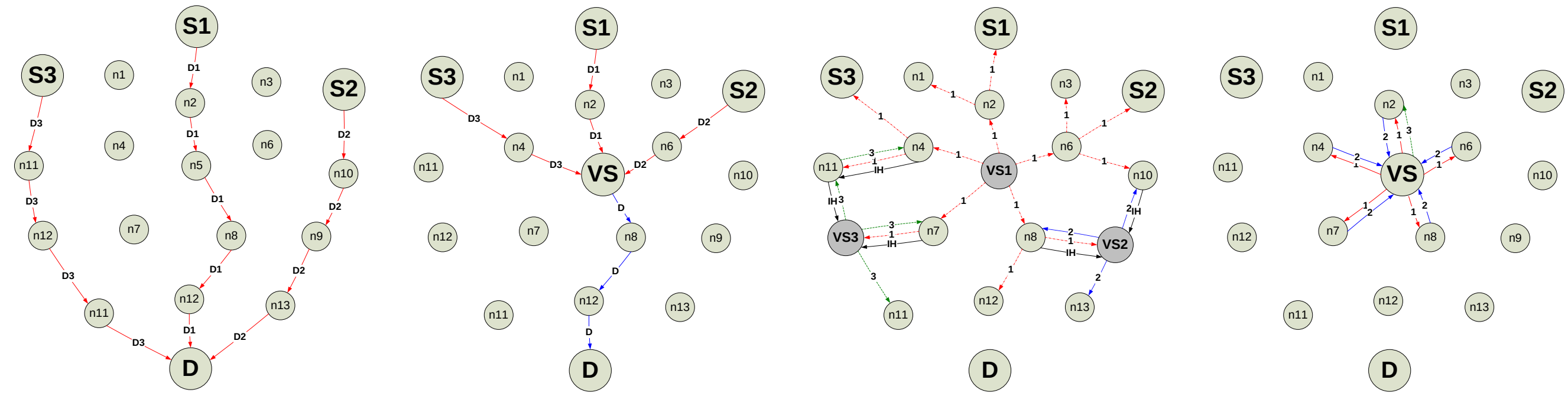

a) Sample Routing Scheme For Directed Diffusion

b) Sample Routing Scheme For Clustered Directed Diffusion

c) Virtual Sink Selection During First Round

d) Virtual Sink Selection During Next Rounds

**Phase III: Virtual Sink Refreshment** – All the traffic generated in the cluster is passed through the VS node. Therefore, the energy of this node will be consumed after a short period of time. So in order to avoid VS failure and to perform load-balancing, after the VS refresh period, VS selects one of its neighbors as the next VS. At first the VS will broadcast a neighbor request message. Then it waits for a short period of time. During this period, VS neighbors will receive this message and send a neighbor reply message which contains energy of each neighbor. So VS can select the neighbor with largest amount of remaining energy. Then it will send a new sink selection message to the selected node. Other important parameters for selecting a node as VS can be its distance from the sink node and other sources but considering complexity of the protocol needed for evaluation of these parameters, it is not a realistic choice. The new VS selection procedure is shown in figure 2.d. In this figure the neighbor request messages are labeled as ‘1’, the neighbor reply messages as ‘2’ and the new sink selection message as ‘3’.

If a new virtual sink cannot find a path toward the sink after a period of time called PATH_EXPLORATION_PERIOD, it will select another neighbor and the VS refreshment steps will be repeated.

**Phase IV: Virtual Sink Expiration** – There are two cases in which the virtual sink is expired. One is when it does not receive any data packet for a period of time called EXPIRATION_PERIOD. The second case is when the remaining energy of VS and all its neighbors become below a threshold called VS_MINIMUM_ENERGY. In this case the VS node will broadcast a VS_DISABLE message in the cluster. The sources after receiving this message will broadcast an exploratory data toward the sink and send the data directly to the sink node. Additionally each node has an expiration timer. This timer is rescheduled by receiving each VS interest message. So if the VS node fails, sources will themselves take the responsibility of sending data toward the sink.

# 3 Multi-path Construction Methods

In this section, we introduce our multi-path construction methods (Nasiri Eghbali and Dehghan 2007). Also we would have a brief evaluation and comparison of these methods.

### Simple Multipath Routing Method

Unlike DD in which the sink node only reinforces the first arriving exploratory data packet and simply drops others, in this method the sink node reinforces all the received exploratory data packets. Also when an old positive reinforcement packet arrives in a node in the path (path node), instead of just dropping the old reinforcement packet (as in DD), it will send a negative reinforcement message to its previous hop. This is done in order to avoid forming none-disjoint paths between sources and sinks. This is used in Ganesan et al. (2001) to increase the resiliency of directed diffusion routing protocol.

As another approach, the path node can also forward the old positive reinforcement message, through the path already traversed by the prior reinforcement message. This way, paths may not be completely disjoint but the number of constructed paths will be fairly increased. These methods are explained in figure 3. The exploratory data (ED) that firstly arrived in each node are labeled as ”e”. In figure 3.a we have depicted the disjoint path creation method. As it is shown, in this method node n5 sends a negative reinforcement message to node n1 after receiving the old P3 reinforcement message.

### Disjoint Proactive Multi-Path Construction Method

This method produces multiple paths proactively by tagging the ED messages in the source node, before forwarding them to the neighbor nodes. We call this tag, multi-path identifier as MP_ID. Basically, this approach acts similarly as the simple disjoint path construction method but when an exploratory data arrives in the sink node, it will be recognized as a new exploratory data (ED) with a new MP_ID or a duplicated one. So the sink reinforces only the first ED packet among those arriving with the same MP_IDs. In this way, we will be assured that created paths are disjoint (the old ED packets are ignored in the base DD algorithm).

This method is shown in the figure 3.b. in which the sink node receives the ED packets with MP_IDs 1 and 2. Firstly the sink receives the ED from node n2 with MP_ID, then receives the ED from node n3 and finally receives the ED from n1. So the sink reinforces only the gradients from node n2 and n3 and ignores the ED received from n1.

The simple and proactive multipath routing algorithms are based on the arrival of ED packets in the sink node with totally distinct path nodes. Unfortunately this rarely happens because of the broadcasting and physical nature of wireless medium, which reduces the probability of arriving multiple different ED packets. The first packet arrives at the sink node, usually traverses all the neighbor nodes and the other ED packets can hardly reach them.

**Figure 3** Comparison Between Simple and Proactive Multi-Path Construction Algorithms

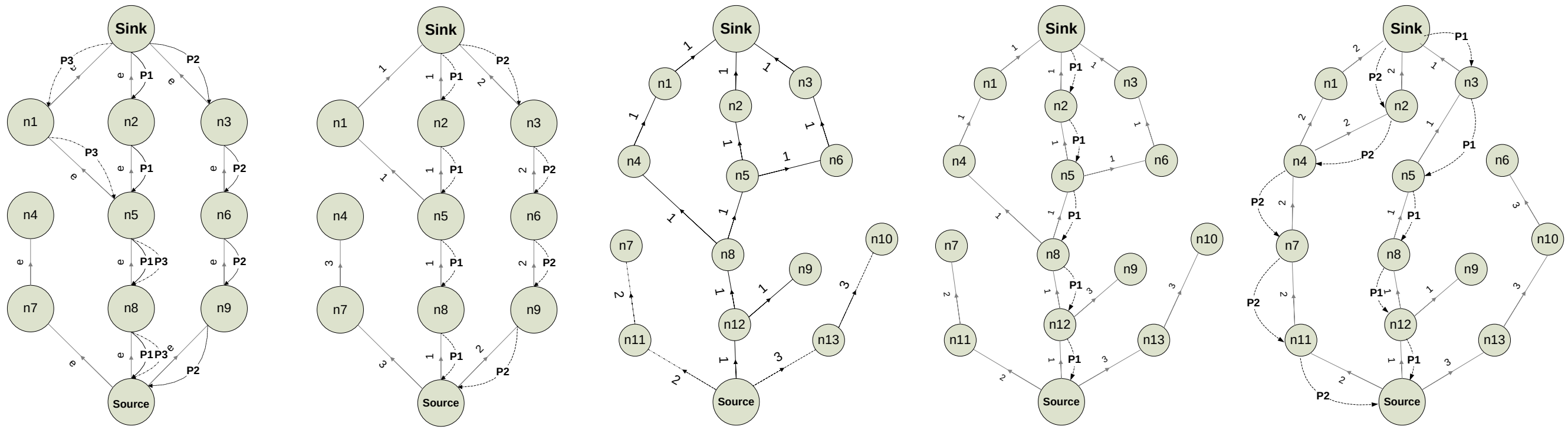

a) Simple Shared Multi-Path Construction

b) Proactive Disjoint Multi-Path Construction

c) Premier Packet Problem

d) Proactive Disjoint Multi-Path Directed Diffusion

e) Disjoint Proactive Multi-Path Method (Using *LFI* Technique)

**Limited Forward Improvement Method**

The functionality of simple multi-path construction methods mostly is related to the topology of the network. The main weakness of these methods relies upon this fact that it rarely can produce multiple disjoint paths. Disjoint routes, starting from the source node, will merge and become a single path after traversing just a few nodes. The main reason behind this problem is the nature of wireless medium and contentions occur during flooding a packet in a wireless network. When the sink node broadcasts an ED packet to its neighbors and they forward the packets to their neighbors, the wireless medium will soon become saturated. In this case the first ED packet that is sent towards the source node (out of the saturated area), can traverse most of the nodes between the source and the sink nodes. We call this event the "premier packet problem" which reduces the probability of arriving multiple distinct ED packets in the sink node. The first packet arrives at the sink node, usually traverses all the neighbor nodes and prevents other ED packets to reach the sink node.

The directed diffusion algorithm makes use of this event for data aggregation but this event prevents the simple and proactive routing methods to make multiple disjoint paths. This event has been depicted in figure 3.c. In this figure the ED packet with MP_ID 1 has reached the sink before others.

Limited Forward Improvement method (LFI) is introduced to solve this problem. In this method, each node selectively forwards the ED packets to first F nearer nodes to the sink. The distance is measured by the time-stamp, saved during the interest message broadcasting (in the first phase of the TPP algorithm). An example is shown in figure 3.d.

The LFI (Limited Forward) improvement method leads to better results and can usually create a notable number of disjoint paths. Another advantage of using this method is reducing the cost of flooding ED packets. Because of its limited forwarding, the number of flooded ED packets will be increased linearly by the size or density of the network. Although in this method, the nodes should send the ED packets to each neighbor instead of broadcasting this method can reduce the total number of communication overhead. In figure 3.e the result of using LFI method has been depicted.

**Braided Proactive Multi-Path Construction Method**

For increasing the number of constructed paths, each node can forward more than just a single ED packet. In this case, the duplicated ED packets with different MP_IDs are not discarded and each node is permitted to forward at most T duplicate packets. We call this parameter, the node forwarding threshold parameter. Therefore, the paths constructed are not necessarily disjoint and can share a number of nodes. Also in this method, gradients are not shared between two different paths and each gradient can only be a member of a single path. As the numbers of forwarded exploratory packets are increased linearly by T value, this method can only be used with LFI (limited forward) improvement method and in this case the imposed routing overhead is almost negligible.

**One and a Half phase Pull Directed Diffusion**

The proposed proactive multipath construction methods, stand somewhere between one-phase-pull (OPP) and two-phase-pull (TPP) algorithms (Heidemann et al. 2003) and can be named as One and a Half phase Pull Directed Diffusion algorithm (OHPP). In the OPP method, the ED packets are omitted and the source node will send data packets to the first neighbor node that has sent the interest packet. In this algorithm we assume the connections between nodes to be bi-directional but this assumption usually is not true. The TPP directed diffusion algorithm has also the problem of large overhead, caused by flooding ED packets through the whole network. In our multipath routing algorithm, the connections need not to be bi-directional and by using LFI improvement techniques, the overhead of ED packets are almost negligible comparing to the TPP algorithm.

**Energy-efficiency and load-balancing**

As mentioned in the introduction, sensor nodes are energy constrained. In this section we present methods for implementing load-balancing mechanisms using the multiple paths, created by the proposed algorithm in the prior sections.

As we know, most of energy utilized in a sensor node is usually consumed for the routing procedure especially in large scale sensor networks where the distance between the source of data and the destination is significant. So it is very important to spread this overhead between different nodes. In this paper the multi-path routing method is used mostly for load-balancing. Although constructing a couple of disjoint or partially-disjoint paths between sink and source will lead to longer life-time of each path and also the connection will be more robust against failures. Therefore, the rate of interest flooding can be reduced which is an expensive operation due to its large communication overhead, although this subject is beyond this work. In this paper we will focus on load-balancing between multiple paths created by the proposed multipath routing algorithms.

A simple approach to perform load-balancing is spreading data packets uniformly between different paths. Then PATH_ALIVE messages can be sent periodically through each path. If a path becomes unusable due to energy consumption of its nodes or other reasons, this path will not be selected anymore.

In this paper we present a more efficient approach in which the path minimum energy (e) and the path length (l) are regarded during the packet spreading process between the alternate paths.

Path minimum energy, e is defined as the energy of node with the least amount of energy between the nodes along the path and the path length is defined as the number of hops in that path. For reinforcing the paths with bigger amounts of e, we consider the probability of sending a packet along a path, directly proportional to its minimum energy. Let $e_p(i)$ be the path minimum energy of path with MP_ID i and $P_{sel}(i)$ be the probably of selecting path i. In this case:

$$P_{sel}(i) \propto e_p(i) \tag{1}$$

Another parameter used is the path length l, as the path length grows, the cost of routing along that path would be increased linearly by the length of it. So we prefer to select the paths with shorter lengths. So:

$$P_{sel}(i) \propto \frac{1}{l_p(i)} \tag{2}$$

By this approach, at first, shorter paths have more chance to forward the data packets but after a period of time, their energy would be decreased. In this situation the longer paths with more energy, would have a better chance of being selected by the source node.

In order to further improve this we introduce two thresholds named as minimum energy threshold (eth) and maximum path length (lth). When the energy of a path reaches blow eth, or its length is more than lth, the probability of its selection by the source node, calculated according to the formulas (1) and (2) multiplied by 0.1. In our simulations, we use the lth with value lmin * β and the eth is selected as emax / α where lmin is the length of the path with the minimum length and emax stands for the e (minimum energy) of the path with the maximum e. (when these parameters are calculated and where are they maintained and how are they updated). So the probability for the source node to select a path between multiple paths is calculated by:

$$P_{sel}(i) = c\ \frac{e(i)}{l(i)} \tag{3}$$

And n is the number of paths constructed by the routing algorithm and α and β are threshold factors. Here we assumed these values to be α = 2 and β = 4 according to our simulation experiments.

$$e(i) = \begin{cases} e_p(i) & e_p(i) > e_{th} \textit{ and } n > o \\ 0.1 * e_p(i) & e_p(i) < e_{th} \end{cases} \quad , e_{th} = \max[e_p(i)] / \alpha$$

$$l(i) = \begin{cases} l_p(i) & l_p(i) < l_{th} \textit{ and } n > o \\ 10 * l_p(i) & l_p(i) > l_{th} \end{cases} \quad , l_{th} = \min[l_p(i)] \times \beta \tag{4}$$

$$c = \frac{1}{\sum_{i=1}^{i=n} \frac{e(i)}{l(i)}} \tag{5}$$

The eth changes dynamically during the routing period. To inform the source node, about the energy of each path, periodic PATH_ALIVE messages are sent along every path. The PATH_ALIVE messages that used for path refreshment are the same as POSITIVE_REINFORCEMENT messages and calculate the minimum energy of the paths. Using this method, the connection time between source and sink can be increased and the number of packets dropped during the routing process can be reduced.

Another way to increase the energy efficiency of routing protocol is using the LFI method that reduces the overhead of flooding ED packets. As we will see in the simulation section, this can increase lifetime of the network significantly.

# 4 EDAP Data Gathering Protocol

As described in former sections, on-demand clustering protocol algorithm aims to reduce the routing overhead by omitting extra exploratory data packets, broadcasted by nearby sources. In this algorithm, nearby sources forward their data towards a node near them which plays the role of the sink node named as virtual sink (VS). Then all data will be aggregated in this node and sent through a single path toward the original sink node. Therefore this node plays the role of single virtual source either.

As highlighted before, using a single path in order to forward all data gathered by sources has the problem of network partitioning. EDAP algorithm uses multipath routing methods presented in section 2 for implementing load-balancing between the VS and the sink node. Disjoint multipath construction method prevents the gathered data to be aggregated but by using on-demand clustering technique, all the data sent toward the sink through multiple paths are already aggregated in the virtual sink node. Figure 1 shows the way data are gathered using EDAP protocol.

# 5 Methodology and Simulation

In this section the methodologies and assumption used for our simulations are described and the simulation results for each proposed algorithm is presented.

### *5.1 Methodology*

Performance evaluation experiments for wireless sensor networks are faced with a number of practical and conceptual difficulties. Here we summarize our main choices for the simulation setup. In this paper on-demand clustering protocol (ODCP) and multipath directed diffusion (MDD) are separately compared with the original directed diffusion (DD) algorithm.

**Protocol Version**

We simulated directed diffusion algorithm available with ns-2 simulator version 2.30. This protocol is implemented for simulator in two versions (Heidemann et al. 2002) named diffusion and diffusion3. Our simulations are based on the diffusion3 protocol which is a complete protocol implementation and allows a more realistic evaluation of the protocol.

**Energy Model**

In the original directed diffusion, the IEEE 802.11 is used for the MAC layer. For comparability we used the same MAC layer and energy model as in Intanagonwiwat et al. (2000) that is the PCM-CIA WLAN card model in ns-2. This card consumes 0.660 W when sending and 0.395 W when receiving. In this paper the transmission range is assumed to be fixed and 200 meters.

**Load Model**

The traditional DD algorithm, floods an interest message, every 30 seconds and exploratory data floods every 50 seconds. We used these predefined values in our simulations. In these simulations, we used the ping application as the network traffic with different packet rates. In ODCP algorithm the rates of 1 packet per second is used and for multipath directed diffusion two different packet rates of 1 and 10 packets per second are used and in EDAP data rate of 0.5 packets per second is used. Also the virtual sink expiration period is assumed to be 60 and 120 seconds.

**Overhead, Drop Percentage and Delay Computation**

In order to compare delay, overhead and drop percentages, between DD and ODCP, these parameters were measured during 100 seconds using variable number of sources. We measured the number of none-data packets, during the connection time and divided it by the number of received data packets to compute overhead. The average delay is also calculated using ping timestamps for each routing algorithm. Drop percentage was measured by dividing number of dropped packets by total number of packets sent.

For comparison between multipath directed diffusion and DD overhead, we measured the number of none-data packets, during receiving 100 data packets in the sink. Our measurement may not be quite precise but it helps us to have a slight comparison between the overhead, imposed by different algorithms.

Also for measuring delay and connection life-time, the source or sources start to send ping data packets towards the sink continually, until the connection is broken, due to path node failures caused by energy depletion. This period is measured and considered as connection life-time. The average delay is also calculated for each routing algorithm. The initial energy of all nodes in the network assumed to be 5 joules. In our simulations, effect of changing VS refresh time using 6 sources and effect of increasing number of sources on connection lifetime was studied. We assumed the initial energy of all nodes in the network to be 5 joules.

**Multi-Path Calculation and Evaluation**

In our approach, for comparison between suggested multi-path routing protocols, protocol is tested in a 10*10 grid (100 nodes), with one sink and one source. Each protocol has been simulated at least five times and the mean value of each measurement has been considered. This iteration was quite necessary, for the random behavior of most proposed algorithms. Then we measured the following parameters: number of paths, total number of share nodes among the paths, average minimum and maximum path length and finally mean path length. We used four densities to highlight the effect of density of nodes on the results of different path construction algorithms. In order to change network density, the minimum distance between two adjacent nodes in the grid is assumed to be 75, 100, 150 and 200 meters.

**Energy Calculations for ODCP**

ODCP algorithm aims to decrease the overhead of routing in network nodes by omitting extra ED packet flooding. For measurement of energy-efficiency of ODCP and comparison between ODCP and DD, we used the scenarios, presented in III.D. Initial energy of all network nodes assumed to be 50 joules. In our scenarios, the sources start to send ping data packets towards the sink continually, during simulation period (250 seconds) and average energy of nodes were measured each 50 seconds.

## *5.2 Simulation Results*

In this section we will show the simulation results, achieved by implementing the scenarios and assumptions, described in this section for ODCP, MDD and ODCP-MP algorithms.

### *5.2.1 On-Demand Clustering Protocol (ODCP) Simulation Results*

**Energy Efficiency**

As it can be figured from figure 4.a, the energy consumption in ODCP is less than DD. This is due to omitting the ED packets for each source node. Also it is obvious that by increasing the number of sources or increasing network size in comparison to cluster size, the efficiency of ODCP will be increased comparing to DD algorithm. Although by using ODCP algorithm, average energy of all network nodes is increased, average energy of cluster nodes will be decreased due to clustering protocol overhead.

**Connection Life-time**

Connection life-time has been shown in figure 4.d. The results show that the connection life-time will be decreased using the ODCP algorithm using less than 4 sources. This decreasing is due to clustering protocol overhead. However lifetime will be improved when the number of sources grows. In these simulations a fixed VS is used during connection lifetime and VS refresh phase has been omitted.

In figure 4.e, effect of changing VS refresh period on connection lifetime has been shown. In this figure VS(R60) and VS(R120) represents using refresh periods of 60 and 120 seconds respectively and VS(base) stands for ODCP without VS refreshment phase. Result show that by decreasing the VS refresh period, although routing overhead will be increased, connection lifetime will be increased significantly. However as it is shown in figure 4.e number of delivered packet does not increase as connection life time. So in this case after 200 seconds, most of sources are disconnected from the sink but at least one of them is connected and therefore connection is not broken until second 430.

**Drop Percentage**

Drop percentage will be increased using ODCP approach because during cluster setup, packets will be dropped (figure 4.c) but this value remains constant and is not sensitive to increasing number of

**Figure 4** On-demand clustering protocol simulation results

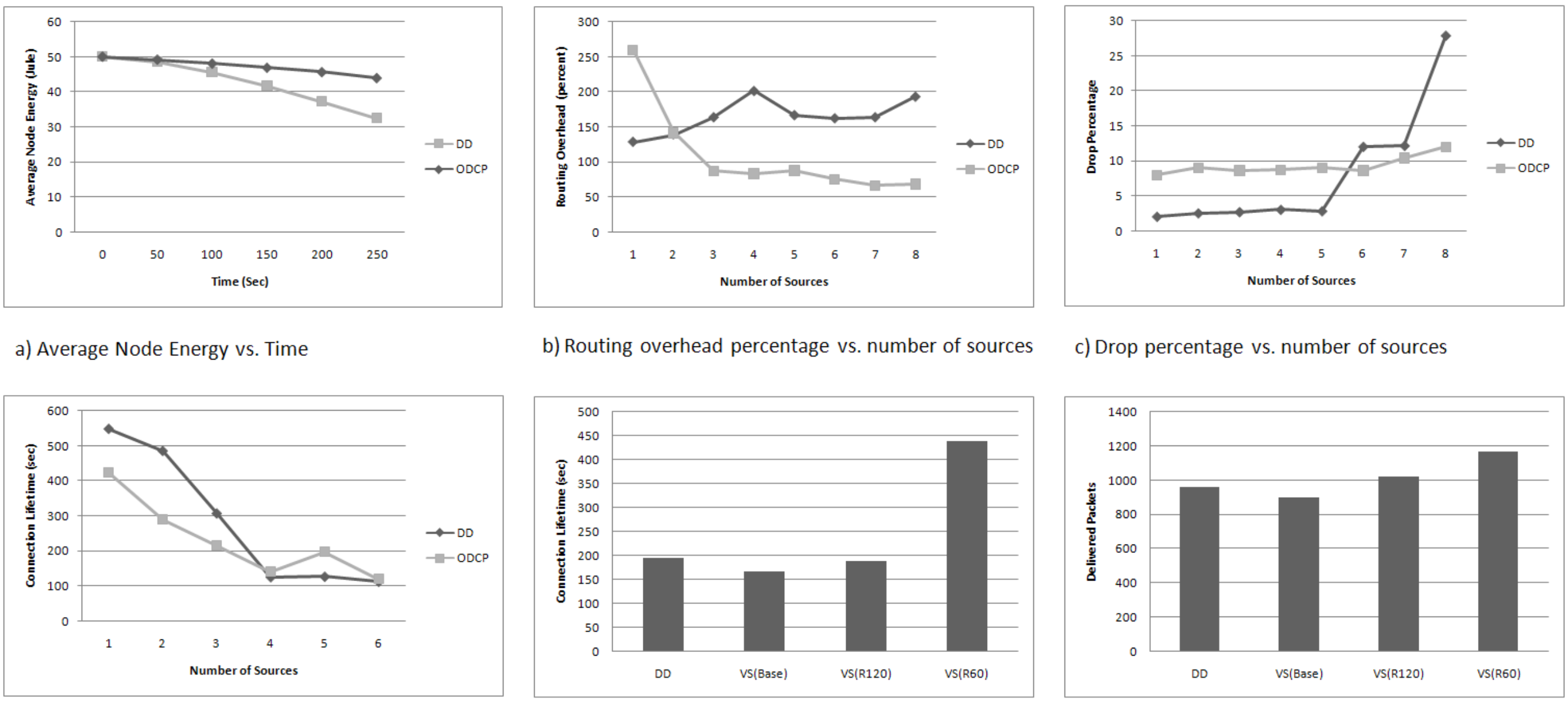

a) Average Node Energy vs. Time

b) Routing overhead percentage vs. number of sources

c) Drop percentage vs. number of sources

d) Connection lifetime vs. number of sources

e) Connection lifetime using different refresh periods

f) Number of delivered Packets using different refresh periods

sources but in DD by increasing number of sources drop percentage will be increased considerably.

### Average Delay

The delay will be decreased using our on-demand clustering protocol. This decreasing is due to lack of network contention produced by flooded exploratory data. However network response time will be decreased because of time needed for cluster setup and VS selection procedure.

### Routing Overhead

The reduction of routing overhead using ODCP comparing to DD has been shown in figure 4.b. This reduction is due to omitting extra ED packets.

### *5.2.2 Multipath Directed Diffusion (MDD) Simulation Results*

### Multipath Construction

In figure 5.a, the number of paths, constructed by different approaches has been shown for different algorithms. Two main results can be extracted from this chart. First the number of paths produced by the different multi-path algorithms grows by increasing the density of the network. Second, among the presented methods, braided proactive multipath routing algorithm

**Figure 5** Multipath routing simulation results

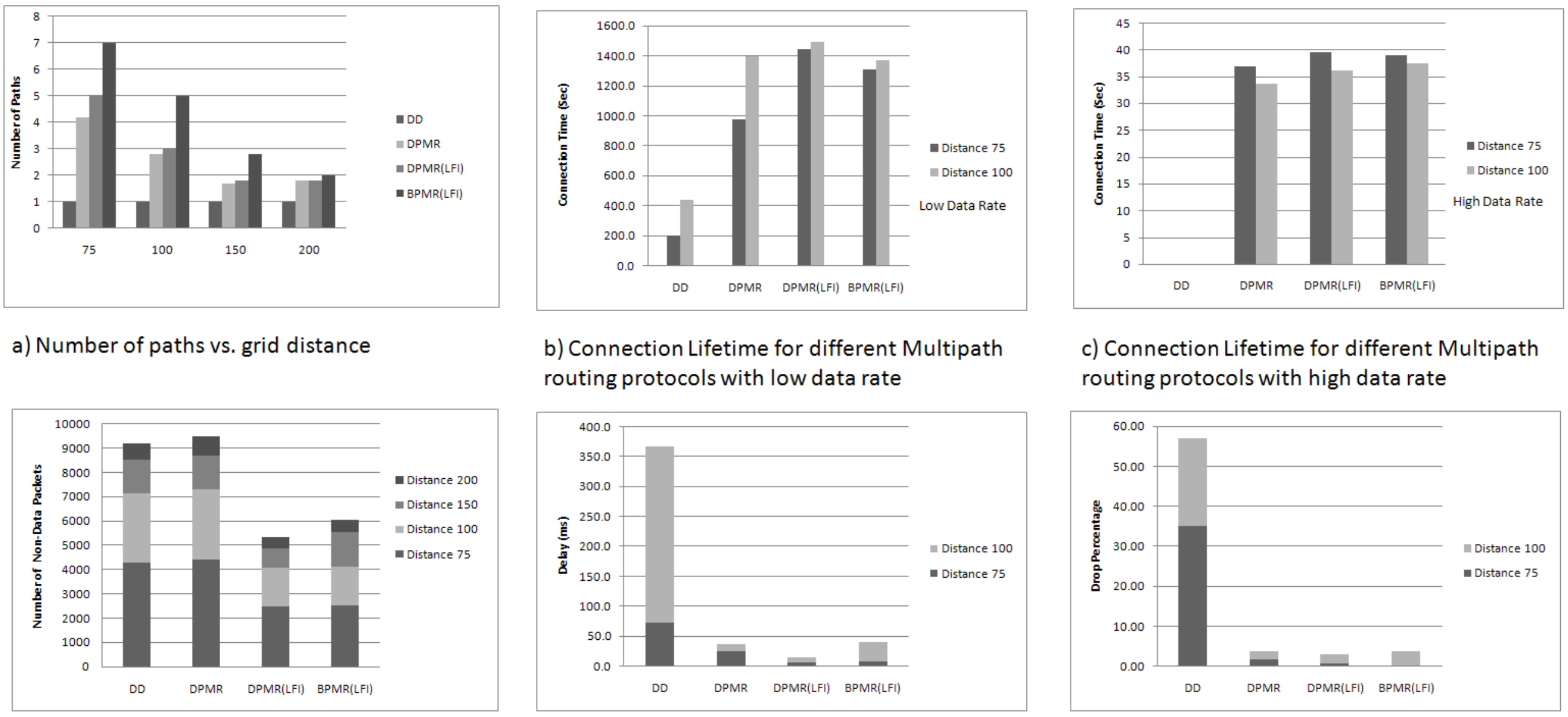

a) Number of paths vs. grid distance

b) Connection Lifetime for different Multipath routing protocols with low data rate

c) Connection Lifetime for different Multipath routing protocols with high data rate

d) Number of none-data packets for different Multipath routing protocols

e) Delay for different Multipath routing protocols with low data rate

f) Drop percentage for different Multipath routing protocols with low data rate

(using LFI improvement method) can produce more paths than the other methods.

**Connection Time**

Using multi-path routing and load-balancing approaches will increase the connection time between source and sink. As shown in figure 5.b and 5.c, multi-path routing has a prominent effect of the connection life-time. The proactive multipath routing algorithm performs much better than the original DD algorithm. This is due to the limited forwarding improvement of this approach which reduced the overhead of exploratory data packet flooding. Also in high data rates (10 packet per second) where the original DD fails, MDD algorithm can deliver packets to destination.

**Overhead Estimation**

The overhead of the directed diffusion will decrease by using LFI method and the proactive routing algorithms show a significant reduction in routing protocol overhead in comparison to DD algorithm. (Figure 5.d)

**Delay Computation and Drop Percentage**

As it is shown in figure 5.e, the multi-path routing methods, will reduce the delay significantly due to omitting the extra ED packets. The drop percentage is also improved significantly using multi-path methods as depicted in figure 5.f.

### *5.2.3 EDAP Simulation Results*

**Energy consumption**

As it is shown in figure 6.a, using EDAP will increase the average energy of network nodes but due to overhead of multipath construction methods, by increasing the number of paths between virtual sink and sink nodes in EDAP protocol, average energy of network nodes will be decreased but better load-balancing will be implemented. In this simulation, 4 sources are used different number of paths are used between virtual sink and the sink node.

**Network Lifetime**

If we define the network lifetime by the period in which at least 50% of network nodes are alive, using EDAP protocol can gain an increase of about 75% in network lifetime (as it can be figured from figure 6.b). Using limited forward improvement method (one and a half phase pull directed diffusion) also increases network lifetime about 25% in our simulations by using 6 sources (figure 6.c) but this gain can be increased in large scaled networks up to 50%.

## 6 Related Works

Classical multi-path routing and clustering protocols has been proposed for both wired and wireless networks. The multi-path routing approaches have been used to perform load balancing and fault tolerance. Alternate path routing schemes in ad-hoc networks have been investigated like TORA (Nasipuri and Das 2000) which provides multipath by maintaining a destination-oriented DAG for each node in the network but the overhead for maintaining the DAG in the network is significant. Multipath extensions to DSR (Park and Corson 1997) support the construction of alternate paths using the source routing mechanism. This works use disjoint path from intermediate nodes on the primary path to enhance resilience. Taheri Javan et al. (2009) proposed a multi-path protocol based on DSR which utilizes common omni-directional antennas, rather than directional ones and transfers data trough multiple zone-disjoint paths simultaneously.

In ad-hoc networks where the energy efficiency is not as restricted as sensor networks and the number of nodes is limited to a few hundred, the multipath routing overhead is usually negligible. Because of data-centric and localized routing algorithms, used in wireless sensor networks, multi-path routing proved to be a complex problem because in such environments, each node can have a local view of the network and no information about the location of the source or the destination is available. Nasiri Eghbali et al. (2009) described the use of multi-sink routing algorithm (MSDD) for implementing load-balancing and increasing the energy efficiency of the routing algorithm in WSN.

In Ganesan et al. (2001) two methods have been proposed for construction of disjoint and braided paths. This work focuses on increasing the resiliency of sensor networks and load-balancing issues have not been considered. We called their disjoint routing protocol as simple multipath routing method. Our proactive multipath routing method in its basic form, behave similarly as their work but their definition used for the braided paths are quite different from ours. The braided paths are defined in their work as the set of paths, each excluding a specific node in the primary path, constructed by the DD algorithm. The Gear algorithm (Yu et al. 2001) relies on geographic information to forward the interest packets in the desired location. Energy efficiency is gained in this method during path selection phase and by reducing the flooding traffic of the interest packets all through the network. But they still use the classic DD algorithm in a more limited area. Our proposed multipath algorithms can be combined easily with this algorithm to produce multiple paths in the interested area. Also in Yu et al.

**Figure 6** EDAP routing simulation results: a) energy usage comparison between DD and DPMR routing method with different number of paths using 6 sources. b) network lifetime comparison between DD and DPMR routing method with different number of paths using 6 sources. c) network lifetime comparison between DPMR and improved DPMR multipath routing methods.

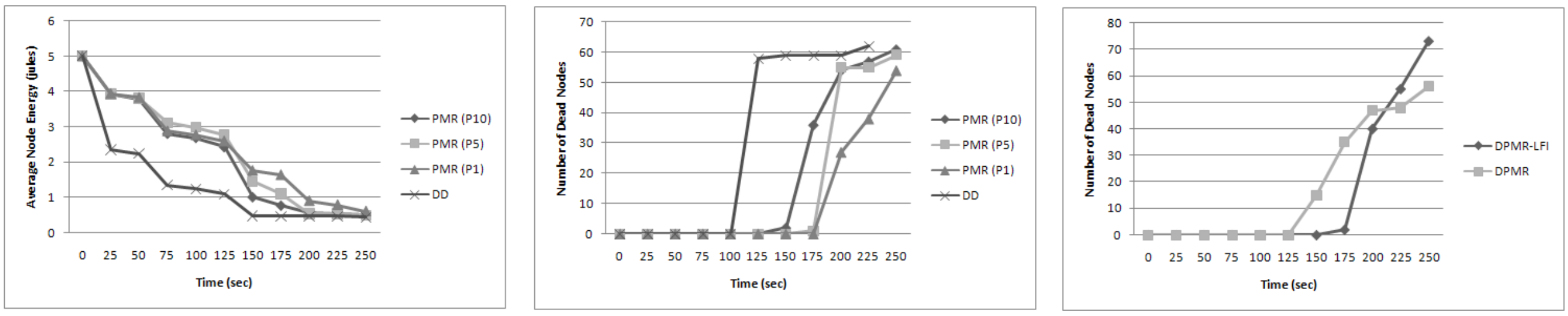

a) Comparision between energy usage of DD and DPMR algorithm with different number of Paths

b) Comparision between number of dead nodes in DD and DPMR algorithm with different number of Paths

c) Comparision between number of dead nodes vs time in simple DPMR and improved DPMR(LFI) methods

(2001) two type of real-time and best effort gradients has been proposed. They used RT gradients to reduce the ETE delay in delay-sensitive applications and BE gradients for performing load-balancing and increasing the energy efficiency of their routing protocol. Although EDDD routing protocol can produce different paths for different kinds of packets, the quality and the quantity of such paths has not been discussed in their work. We also used the idea presented in Braginsky and Estrin (2002) for RFI improvement method.

Heinzelman et al. (2000) proposed LEACH protocol which is the base idea used in many other clustering schemes. This protocol consists of rounds in which cluster heads are selected probabilistically in the beginning. The cluster head gathers cluster data using TDMA scheduling and send aggregated data directly to sink. Improvements to Heinzelman et al. (2000) are proposed in TEEN (Manjeshwar and Agrawal 2001) and PEGASIS (Lindesy and Raghavendra 2002). In TEEN, soft and hard threshold s has been used in each node for sending data in order to decrease the data transmission rate. PEGASIS is a chain-based power efficient protocol that forms the chain toward the base-station greedily using geographical information of each node. The chain leader aggregates the data and sends them toward the sink. Cluster heads are not distributed evenly in LEACH. MECH tries to address this problem by gathering local neighbor information to assign cluster heads. Cluster heads in this protocol routes data among cluster heads to reach the base station instead of sending them directly to the sink.

# 7 Conclusion

In this paper Efficient Data-gathering protocol (EDAP) for directed diffusion algorithm is proposed for situations where an event can trigger more than a single sensor node. We used a virtual sink node near the sources. This node gathers information sent by nearby sources and undertakes responsibility of sending them toward the sink. This way, extra exploratory data packets will be omitted and data from different sources will be aggregated as soon as possible.

Although improvements gained by early-aggregation have not been considered in our simulations, results show that routing overhead will be decreased significantly. This improvement is gained through omitting the extra exploratory data packets flooded by each source. Also connection life-time between sources and sink will be increased using proper virtual sink refresh periods.

This paper also describes the use of multipath routing method for implementing load-balancing and increasing the energy efficiency of the routing algorithm. In this work proactive method were proposed for constructing multiple paths Limited Forward approach is presented to improve the efficiency of this algorithm. The LFI method also decreases the number of exploratory data packets flooded in the network, significantly and reduces the overhead of two-phase pull algorithm. Simulation results also show that using multipath routing algorithm leads to longer connection lifetimes.

Each of these algorithms has its own weak points. When using on-demand clustering protocol, all paths from different sources will merge after a few hops and all of them are forwarded by virtual sink node toward the sink, using a single path. As mentioned earlier this may cause the network to be partitioned. Also proactive multipath routing algorithm constructs disjoint paths between sources and sink nodes and prevents aggregation. EDAP is introduced to solve these problems by using the benefits of both algorithms and simulation results show that using EDAP protocol can lead to 75% improvement in network lifetime in comparison to directed diffusion. However using multipath routing technique in the EDAP algorithm may not increase the connection life-time considerably. The reason behind this is that VS node will die before other nodes and breaks the connection but in real world applications usually the nodes near the sink node will die before the other nodes and by using EDAP protocol, the life-time of these nodes will be increased by performing load-balancing techniques.